\documentclass[12pt]{article}
\usepackage{graphicx}
\usepackage{dcolumn}
\usepackage{bm}
\usepackage{amsmath, amssymb, amsfonts, amsthm}
\usepackage{cite}
\usepackage{epsfig}
\usepackage{textcomp}
\usepackage{tikz-cd} 
\usepackage{ dsfont }

\def\a{\alpha}
\def\b{\beta}
\def\g{\gamma}

\def\e{\epsilon}

\def\f{\phi}

\def\m{\mu}
\def\n{\nu}
\def\r{\rho}
\def\s{\sigma}

\def\o{\omega}

\def\be{\begin{equation}}
\def\ee{\end{equation}}
\def\ba{\begin{eqnarray}}
\def\ea{\end{eqnarray}}

\def\bdm{\begin{displaymath}}
\def\edm{\end{displaymath}}

\def\bq{\begin{quote}}
	\def\eq{\end{quote}}

 at 10truept

\newcommand{\bea}{\begin{eqnarray}}
\newcommand{\eea}{\end{eqnarray}}

\newcommand{\bi}{\begin{itemize}}
	\newcommand{\ei}{\end{itemize}}

\newcommand{\beq}{\begin{equation}}
\newcommand{\eeq}{\end{equation}}
\newcommand{\beqa}{\begin{eqnarray}}
\newcommand{\eeqa}{\end{eqnarray}}

\def\ltap{\ \raise.3ex\hbox{$<$\kern-.75em\lower1ex\hbox{$\sim$}}\ }
\def\gtap{\ \raise.3ex\hbox{$>$\kern-.75em\lower1ex\hbox{$\sim$}}\ }
\def\gl{\ \raise.5ex\hbox{$>$}\kern-.8em\lower.5ex\hbox{$<$}\ }
\def\roughly#1{\raise.3ex\hbox{$#1$\kern-.75em\lower1ex\hbox{$\sim$}}}

\begin{document}

		\thispagestyle{empty}
		\begin{flushright}
			September 2017\\
		\end{flushright}
		\vspace*{.2cm}
		\begin{center}
			{\Large \bf  Spontaneously Broken Spacetime Symmetries\\ \vspace{0.25cm} and the Role of Inessential Goldstones}

			\vspace*{.7cm} {\large  Remko Klein\footnote{\tt
					remko.klein@rug.nl}, Diederik Roest\footnote{\tt d.roest@rug.nl
				} and  David Stefanyszyn\footnote{\tt
				d.stefanyszyn@rug.nl}}
		
		\vspace{.3cm} {\em Van Swinderen Institute for Particle Physics and Gravity, University of Groningen, Nijenborgh 4, 9747 AG Groningen, The Netherlands}\\

		\vspace{1cm} ABSTRACT 
	\end{center}

In contrast to internal symmetries, there is no general proof that the coset construction for spontaneously broken spacetime symmetries leads to universal dynamics. One key difference lies in the role of Goldstone bosons, which for spacetime symmetries includes a subset which are inessential for the {non-linear realisation} and hence can be eliminated. In this paper we address two important issues that arise when eliminating inessential Goldstones.

The first concerns the elimination itself, which is often performed by imposing so-called inverse Higgs constraints. Contrary to claims in the literature, there are a series of conditions on the structure constants which must be satisfied to employ the inverse Higgs phenomenon, and we discuss which parametrisation of the coset element is the most effective in this regard. We also consider generalisations of the standard inverse Higgs constraints, which can include integrating out inessential Goldstones at low energies, and prove that under certain assumptions these give rise to identical effective field theories for the essential Goldstones.

Secondly, we consider mappings between non-linear realisations that differ both in the coset element and the algebra basis. While these can always be related to each other by a point transformation, remarkably, the inverse Higgs constraints are not necessarily mapped onto each other under this transformation. We discuss the physical implications of this non-mapping, with a particular emphasis on the coset space corresponding to the spontaneous breaking of the Anti-De Sitter isometries by a Minkowski probe brane.

	\vfill \setcounter{page}{0} \setcounter{footnote}{0}
	\newpage
	
	\tableofcontents
	
	\section{Introduction} \label{intro}

The spontaneous breaking of symmetries is of critical importance in many areas of physics. For an internal symmetry group $G$ which is spontaneously broken to a subgroup $H$, the tools to construct the non-linear realisation of the group $G$ were developed by Callan, Coleman, Wess and Zumino (CCWZ) in the late 1960's \cite{internal1,internal2}. In this coset construction there is a single Goldstone boson for each broken generator and the dynamics of the Goldstones is dictated by the coset space $G/H$. Moreover, for compact, semi-simple groups, it has been proven that all non-linear realisations of such a spontaneously broken symmetry are related by invertible field redefinitions, and as a consequence can be derived from the coset construction. This guarantees the universality of all corresponding observables. 

The generalisation of the coset construction of CCWZ to spontaneously broken spacetime symmetries came a few years later \cite{spacetime1,spacetime2} and has been used extensively in the context of constructing and understanding effective field theories used for model building in cosmology and gravity. Two notable examples are the scalar sector of the $d$-dimensional DBI Lagrangian which non-linearly realises the $(d+1)$-dimensional Poincar\'{e} group, see e.g. \cite{wess}, and the Volkov-Akulov Lagrangian which non-linearly realises supersymmetry with a single fermion \cite{VA}. Both of these theories, and their higher order corrections, can be derived using the coset construction. Complimentary methods include the study of hypersurfaces fluctuating in transverse directions, e.g. \cite{sundrum,reunited,VAbrane,higher}, and the study of soft limits of general scattering amplitudes, e.g. \cite{soft1,soft2,soft3}. See also \cite{condensed} for a discussion on spontaneous breaking of spacetime symmetries in condensed matter systems, \cite{super1,super2} for a discussion on the coset construction for superfluids etc and \cite{cosmology1,cosmology2,cosmology3,cosmology4}  for more examples related to cosmology and gravity.

However, the coset construction for spacetime symmetries involves added subtleties compared to the case of internal symmetries. Chief amongst these is a distinction between the Goldstone modes corresponding to all broken generators: not all Goldstones are essential in order to non-linearly realise the broken symmetry group. Instead, consistent non-linear realisations exist where the number of Goldstone fields is less than the number of broken generators, with only the essential Goldstones enjoying the special symmetry protection of the non-linear realisation. In order to dispense of the inessential modes one can impose inverse Higgs constraints\cite{spacetime2}\footnote{For discussions on the physical origin of inverse Higgs constraints we refer the reader to \cite{low,gapped1,gapped2}.}. A very clear example of this is the conformal group in four dimensions spontaneously broken to its four dimensional Poincar\'{e} subgroup \cite{conformal}. There are five broken generators yet a consistent non-linear realisation exists with a single Goldstone field, the dilaton, while the vector of the broken special conformal transformations is redundant.

The process of elimination of the inessential Goldstone bosons complicates the relation between different non-linear realisations. In particular, it is not known whether different coset constructions that non-linearly realise the same symmetry are equivalent.  In order to make progress in this direction, this paper deals with two crucial aspects pertaining to the elimination of the inessential Goldstone modes. 

The first part is focused on the intricate link between the existence of inverse Higgs constraints and the parametrisation of the coset element. After reviewing the most important aspects of the coset construction for both internal and spacetime symmetry breaking in section \ref{coset}, in section \ref{inverseHiggs} we present the conditions on the structure constants which must be satisfied in order to employ the inverse Higgs phenomenon. Here we focus on standard inverse Higgs constraints i.e.~where the inessential Goldstones are eliminated algebraically by setting a covariant derivative to zero. Contrary to what is often stated in the literature, there are a \textit{series} of conditions which need to be met rather than a single one. Notably, the standard parametrisation considered in the original work \cite{spacetime1,spacetime2} is not the optimum one in this regard, and already fails for the very simple case of spontaneous breaking of the $d$-dimensional Poincar\'{e} group down to its $(d-1)$-dimensional subgroup. Instead, the parametrisation which requires the least stringent conditions for the existence of an inverse Higgs constraint  involves a further splitting of the broken generators compared to the standard parametrisation. 

In section \ref{inverseHiggs} we also discuss the possibility of imposing ``generalised" inverse Higgs constraints. These constraints again allow one to eliminate inessential Goldstone fields but they do not follow from the usual inverse Higgs phenomenon as outlined in \cite{spacetime2}. An example would be an equation of motion either where an inessential Goldstone is an auxilliary field and can be eliminated algebraically in comparison to the standard inverse Higgs constraint or where an inessential Goldstone is integrated out at low energies. The latter is possible since inessential Goldstones can acquire a mass consistent with all the symmetries. In some cases equations of motion are equivalent to the standard inverse Higgs constraints \cite{mcarthur} but this is not always the case. In any case we show that as long as the series of inverse Higgs conditions are met, the effective field theory constructed from only the essential Goldstone fields is the same, up to variations in coupling constants which are not fixed by symmetry, regardless of how one chooses to eliminate the inessential Goldstones. This equivalence has been mentioned in the literature, e.g.~\cite{brauner,wheel}, but to our knowledge this is the first time it has been shown to be true.

The second part of this paper is focused on how differerent non-linear realisations of a broken symmetry group are related to each other. We investigate the relations between coset constructions employing  different parametrisations of the coset element as well as algebra bases in section \ref{mapping}. Prior to imposing inverse Higgs constraints, the relationship between the different non-linear realisations is  straightforward and involves transformations between the coset coordinates, which for spontaneously broken spacetime symmetries includes the spacetime coordinates and the fields. These are known as \textit{point transformations}, and are the natural generalisation of field redefinitions in the internal case. However, the construction of possible transformations becomes much more complicated after we impose inverse Higgs constraints, since the constraints are not necessarily mapped onto each other under point transformations. If they are, the point transformation on the coset coordinates induces a transformation involving the spacetime coordinates, the essential Goldstone fields and their derivatives; these are so-called \textit{contact transformations} or \textit{extended contact transformations}. If the inverse Higgs constraints are not mapped onto each other, the situation is less clear; there can still be extended contact transformations relating the two coset constructions, but they do not follow from the point transformation.

Allowing for changes in the algebra basis may seem like a unnecessary complication, but different bases can have different physical motivation. For example, consider the spontaneous breaking of the $d$-dimensional conformal group $SO(d,2)$ by a $n$-dimensional Minkowski probe brane embedded in $(d+1)$-dimensional Anti-De Sitter (AdS) space. There are two natural bases for the conformal algebra: the standard conformal basis and the AdS basis. The AdS basis is of interest since the resulting non-linear realisation matches the one derived from the usual probe brane construction using the induced metric and its derivatives. To relate this non-linear realisation to the one derived using the coset construction and the standard conformal basis requires exactly the type of transformations we are considering. Interestingly, for $d=n$ (codimension one) both inverse Higgs constraints are mapped onto each other, thus establishing a contact transformation relating the two non-linear realisations \cite{equivalence,mapping}. This transformation reduces to that of the galileon duality\cite{duality,duality1} after taking the appropriate contractions\footnote{The galileon duality transformation can also be extracted more straightforwardly by considering the coset construction for spontaneous breaking of the Galileon group \cite{duality1,duality2}.}. However, as we will discuss in detail in section \ref{conformal}, in higher codimensions with $d>n$, the inverse Higgs constraints of both bases are not mapped onto each other. As a consequence, it is unclear if the equivalence is maintained. In this sense, different algebra bases are a useful way of examining the universality of non-linear realisations of spacetime symmetries.

We end with a conclusion and outlook with particular attention paid to the question of universality for spontaneously broken spacetime symmetries.

\bigskip

\noindent {\bf Notation:} Unless otherwise stated, throughout we denote an arbitrary generator of the group $G$ using indices $I,J,\ldots$, a broken generator using $A,B,\ldots$, an unbroken generator using $i,j,\ldots$ and the spacetime coordinates using $\m,\n,\ldots$. When we discuss the inverse Higgs phenomenon we will assume that $A$ is reducible under the subgroup $H$ and hence splits into multiple irreps, for which we will  use $a,b,\ldots$ for essential and $m,n,\ldots$ for inessential Goldstones.

\section{Coset construction} \label{coset}

In this section we review the coset construction as a tool for constructing non-linear realisations. We begin with the case where the non-linearly realised symmetry is an internal one i.e. the generators commute with those of the Poincar\'{e} group, then we move onto spontaneously broken spacetime symmetries. Readers familiar with the coset construction can jump to section \ref{inverseHiggs}, where a discussion of the inverse Higgs phenomenon with various clarifications which to our knowledge are absent from the literature can be found.

\subsection{Internal symmetries}

Consider a group $G$ with subgroup $H$, where the broken generators  of $G/H$ are denoted by $T_A$ and the unbroken ones of $H$ by $T_i$. For the coset construction to be applicable, one must assume that the generators $T_A$ form a (reducible) representation of the subgroup $H$. In other words we have the following commutators
\begin{align} \label{internalcom}
[T_i,T_j] = f_{ij}^k T_k, ~~~~~~~ [T_A,T_i] = f_{Ai}^B T_B, ~~~~~~~ [T_A,T_B] = f_{AB}^I T_I.
\end{align} 
The aim of the game is to derive the building blocks used to construct Lagrangians which when made manifestly $H$-invariant are automatically $G$-invariant. We note that the coset construction is not restrictive inasmuch as we can take the extreme cases where $H = \mathds{1}$ or $H = G$.

To construct the non-linear realisation consider an element $g$ from the group $G$. Locally one can parametrise this group element in terms of the generators of $G$ as
\begin{align}
g &= e^{\f^A T_A}e^{\f^i T_i}  \,,
\end{align}
which is of course not a unique choice. At the heart of the coset construction is the coset space $G/H$ which is the set of equivalence classes of $G$ under right multiplication of $H$ and one can parametrise an element of the coset as
\begin{align} 
\g (\f) &= e^{\f^A T_A} \,, \label{coset-element}
\end{align}
which is again not a unique choice but merely a standard one, see section 4. From this coset element we can define a non-linear realisation of the group $G$ on the fields $\f^A$, the coset coordinates corresponding to the generators $T_A$, by considering the multiplication 
\begin{align} \label{action}
g\g(\f) h^{-1}(\f,g) \equiv \g(\f') = e^{\f'^A T_A} \,, 
\end{align}
where we used a $H$ transformation from the right to put the coset representative in the specified form since in general multiplication by any element of $G$ does not preserve this choice. Now this action on the coset representative defines a non-linear realisation of $G$ on the coordinates $\f^{A}$ as
\begin{align}  \label{goldstonetransformation}
g\cdot \f^{A} \equiv \f'^{A}(\f,g).
\end{align} 
Next consider some other fields $\psi$ which transform under some linear representation of $H$ but not under the full group $G$. Using the coset coordinates, the linear action of $H$ can be extended to a non-linear realisation of $G$ via
\begin{align}
\psi'^{\a}= D^{\a}_{\b}( h)\psi^{\b} \,, \label{mattertransformation}
\end{align}
where the definition of $h(\f,g)$ follows from \eqref{action}.
Together, the transformation laws (\ref{goldstonetransformation}) and (\ref{mattertransformation}) define a consistent non-linear realisation on $\f^{A}$ and $\psi^{\a}$. As we noted in the introduction, the power of this formalism is that \textit{any} non-linear realisation of a compact, semi-simple internal symmetry can be put into this form by doing a suitable, locally invertible, field redefinition. Such universal statements are not proven for non-compact and/or non-semi-simple groups; however, in this case at least all possible coset constructions (with e.g.~different coset elements \eqref{coset-element}) are related to each other by field redefinitions, see Section 4.

One can now construct Lagrangians which are invariant under these transformations. The trick here is to construct objects which transform covariantly, similar to e.g.~$\psi^{\a}$. These objects can be extracted from the Maurer-Cartan form $\g^{-1}d\g$ which is part of the Lie algebra of $G$ and can therefore be decomposed with respect to the generators as
\begin{align} 
\g^{-1}d\g &= \o^A T_A + \o^i T_i = ((\o^A)_\m T_A + (\o^i)_\m T_i)dx^\m \,,
\end{align}
where the Maurer-Cartan components $\o^A$, $\o^i$ are functions of the coset coordinates. Given the transformation of $\g$ under the action of $G$ we have
\begin{align} 
g \cdot (\g^{-1} d\g) = h (\g^{-1} d\g - h^{-1}dh) h^{-1} \,,
\end{align} 
or in terms of the components
\begin{align} 
g\cdot (\o^A)_\m & = D(h)^A_B (\o^B)_\m \,, \qquad
g\cdot (\o^i)_\m  = D(h)^i_j (\o^j)_\m - D(h)^i_j (h^{-1}\partial_\m h)^j  \,,
\end{align} 
where we used that $H$ is a subgroup and that the broken generators $T_A$ form representations under $H$.

From these transformations we see that the components of the Maurer-Cartan form corresponding to the broken generators transform covariantly as desired. The components corresponding to the unbroken generators do not transform covariantly; instead, they provide the connection terms that one needs to build covariant derivatives for the fields $\psi^{\a}$ and higher order derivatives for the Goldstones. To see this first note that the ordinary derivative does not transform covariantly since
\begin{align} 
g\cdot \partial_\m \psi^\a &= \partial_\m(D(h)^\a_\b \psi^\b) = D(h)^\a_\b \partial_\m\psi^\b  +  \partial_\m (D(h)^\a_\b)\psi^\b  \,, \notag \\
&= D(h)^\a_\b (\partial_\m\psi^\b + (D(h)^{-1}\partial_\m D(h))^\b_\g \psi^\g)  \,,
\end{align}  
but this can be compensated for by introducing the following covariant derivative
\begin{align}
\nabla_\m \psi^\a &= \partial_\m \psi^\a + (\o^i)_\m (T_i)^\a_\b \psi^\b \,, \qquad
g \cdot \nabla_\m \psi^\a = D(h)^\a_\b (\nabla_\m\psi^\b).
\end{align}
We now have a set of covariantly transforming objects, including $\psi^{\a}$, $\nabla_\m$ and $(\o^A)_\m$. Any Lagrangian built from these that is invariant under the linearly realised $H$ will be invariant under the non-linearly realised $G$.

In addition to invariant Lagrangians one can also consider Lagrangians that shift by a total derivative. In this case the $d$-form $Ld^d x$ shifts by an exact $d$-form. As a consequence its exterior derivative is an invariant  $(d+1)$-form. Thus by constructing all invariant exact $(d+1)$-forms, $\a = d\b$, using the covariant building blocks of the coset construction, one can find all $d$-forms $\b$ which are invariant either exactly or up to a total derivative. These $\b$'s can be used to construct Lagrangians and those that shift by a total derivative are Wess-Zumino terms \cite{WZ,witten}.

\subsection{Spacetime symmetries}

Now we consider the case where $G$ and $H$ are no longer purely internal symmetries, but also contain spacetime symmetries. We assume that the subgroup $H$ contains the Lorentz generators. In addition to the commutators considered above (\ref{internalcom}), we assume that the translations $P_\m$ form a representation of $H$. Therefore the commutators are
\begin{align} 
&[T_i,T_j] = f_{ij}^k T_k, ~~~~~~~  [T_A,T_i] = f_{Ai}^B T_B, ~~~~~~  [T_A,T_B] = f_{AB}^I T_I,  \nonumber \\  & [P_\m,T_i] = f_{\m i}^\n  P_\n , ~~~~~ [P_\m,T_A] = f_{\m A}^I T_I.
\end{align} 
A key difference here compared to internal symmetries is that the generators of translations should be included in the coset element, with their coefficient being the spacetime coordinates, since translations act non-linearly on the spacetime coordinates: $x^{\m} \rightarrow x^{\m} + \epsilon^{\m}$. Therefore spacetime coordinates should be interpreted as the Goldstone modes for translations.

Again we can parametrise a group element of $G$ in several ways but for now let us choose the following standard parametrisation
\begin{align} 
g &=  e^{x^\m P_\m}e^{\f^A T_A}e^{\f^i T_i} \,,
\end{align} 
where the $T_i$ contain unbroken internal as well as spacetime symmetries, and the $T_A$ are the broken internal and spacetime symmetries. Again we consider the coset $G/H$ whose standard parametrisation is \cite{spacetime1,spacetime2}
\begin{align}  
\g (x,\f) &=  e^{x^\m P_\m}e^{\f^A T_A} \,,
\end{align} 
and we can define a consistent non-linear realisation on the fields as follows
\begin{align}
g \g (x,\f) h^{-1}(x,\f,g) \equiv \g(x',\f'(x')), \qquad  g \cdot x  \equiv x', \qquad g \cdot \f(x) \equiv \f'(x')  \,.
\end{align}
Likewise a non-linear realisation on fields transforming under some linear representation of $H$ can be defined as
\begin{align}
\psi'^{\a}(x') = D^{\a}_{\b}(h)\psi^{\b}(x).
\end{align}
In order to construct invariant Lagrangians we again use the Maurer-Cartan form which now has the following structure
\begin{align} 
\g^{-1}d\g = \o^\m P_\m + \o^A T_A + \o^i T_i \,,
\end{align} 
and transformation properties 
\begin{align}
g\cdot (\o^\m)_\n  & dx^\n = D(h)^\m_\n (\o^\n)_\r dx^\r \notag, \qquad g\cdot (\o^A)_\n dx^\n = D(h)^A_B (\o^B)_\r dx^\r \notag \\
&g\cdot (\o^i)_\n dx^\n = D(h)^i_j (\o^j)_\r dx^\r - D(h)^{i}_{j} (h^{-1} \partial_{\m} h)^{j} dx^{\m},
\end{align}
i.e. the components $(\o^I)_\m$ do not transform covariantly and we must use the $\o^I$ to build invariant Lagrangians since now the coordinates transform. Also, since it is the object $(\o^\m)_\n dx^\n$ that has nice transformation properties rather than the $dx^\m$ themselves, one interprets the components $(\o^\m)_\n$ as vielbeins
\begin{align} 
e^\m{}_\n \equiv (\o^\m)_\n \,,
\end{align}  
enabling one to define a metric and corresponding invariant measure as follows
\begin{align} 
g_{\m\n} = e^\r{}_\m e^\s{}_\n \eta_{\r\s}, \qquad \sqrt{-g}d^4 x = \e_{\m\n\r\s}\o^\m \wedge \o^\n \wedge \o^\r \wedge \o^\s.
\end{align} 
We can also define a covariant derivative of the fields, which has the desired covariant transformation properties, by using the Maurer-Cartan components along the directions of the broken generators as
\begin{align} 
\nabla_\m \f^A &= (e^{-1})_\m^\n (\o^A)_\n \,, \qquad 
g\cdot \nabla_\m \f^A = D(h)^A_B D(h)_\m^\n \nabla_\n \f^B,
\end{align} 
and similarly we can define the covariant derivative of the matter fields $\psi^{\a}$ using the components along the directions of the unbroken generators as
\begin{align} 
\nabla_\m \psi^\a &=  (e^{-1})_\m^\n (\partial_\n \psi^\a + (\o^i)_\n (T_i)^\a_\b \psi^\b ) \,, \quad
g\cdot \nabla_\m \psi^\a = D(h)^\a_\b D(h)_\m^\n \nabla_\n \psi^\b.
\end{align} 
Similar to the internal symmetry case, one can now construct $H$-invariant Lagrangians out of the objects $\nabla_\m \f^A$, $\psi^\a$ and $\nabla_\m$ and multiply them with the invariant measure $\sqrt{-g}d^d x$ in order to build $G$-invariant actions. Alternatively, one can construct $H$-invariant $d$-forms out of the covariantly transforming objects $\o^\m$, $\o^A$, $\nabla_\m$, $\psi^\a$ to yield an invariant action.

Again one can construct Lagrangians invariant up to a total derivative by adding  Wess-Zumino terms. A well known example of Wess-Zumino terms for spacetime symmetries is Galileons \cite{wess}.

\section{Eliminating inessential Goldstone modes} \label{inverseHiggs}

As we discussed in the introduction, for spontaneously broken spacetime symmetries one does not necessarily need a Goldstone field for every broken generator. Rather there is some reduced set of Goldstones corresponding to a restricted set of broken generators which can still non-linearly realise the broken symmetry. In this section we discuss how one can eliminate the inessential Goldstones. For clarity we focus on cases where there are two Goldstone fields; one essential and one inessential and therefore a single inverse Higgs constraint but our results can be easily extended to more complicated cases too.

\subsection{Standard inverse Higgs constraints}

The main message we wish to convey in this subsection is that $i)$ the existence of inverse Higgs constraints is heavily dependent on the parametrisation of the coset element and $ii)$ the optimum parametrisation in this regard is not the standard one (\ref{standardcoset}) as used in \cite{spacetime1,spacetime2} but rather a parametrisation with further splitting of the broken generators (\ref{splitcoset}).

Once we have chosen a parametrisation for the coset element we can calculate all objects of interest with regards to the non-linear realisation as explained in section \ref{coset}. In terms of eliminating inessential Goldstone fields the object of most interest is the covariant derivative which in terms of the Maurer-Cartan components is given by
\begin{align} 
\nabla_\m\f^A &= (e^{-1})_\m^\n (\o^A)_\n .
\end{align}  
Now we assume that $A$ is reducible under $H$ and hence splits in multiple irreps. Let us distinguish between two, namely, $a$ and $m$. Concentrating on the covariant derivative for the $\phi^{a}$ field we have
\begin{align}  \label{covderiv}
\nabla_\m\f^a &= (e^{-1})_\m^\n (\o^a)_\n,
\end{align}  
which can be expressed in terms of structure constants once we choose a parametrisation for the coset element. The idea of the inverse Higgs phenomenon, as outlined in \cite{spacetime2}, is to use this covariant derivative to \textit{algebraically} solve for $\f^m$ in terms of $\f^a$ and $\partial_\m \f^a$. Assuming $\m \times a \supset m$, it is often stated in the literature that if 
\begin{equation} \label{structureproject}
f_{\m m}^{a}|^{n} \neq 0,
\end{equation}
i.e. there exists a non-zero component of the structure constant $f_{\m m}^{a}$ once we project $\m \times a$ on $n$, one can solve for $\f^{m}$ in terms of $\f^a$ and $\partial_\m \f^a$ by setting
\begin{align}  \label{covderivproject}
\nabla_\m\f^a |^{n} = c \f^{n} + \partial_{\m} \f^{a} |^{n} + \ldots
\end{align} 
to zero. This is because (\ref{structureproject}) ensures that $\f^{m}$ appears linearly. However, in general (\ref{covderivproject}) contains $\partial_{\m} \f^{m}$ terms which \textit{might} restrict one from solving for $\f^{m}$ algebraically. In this sense (\ref{structureproject}) is merely a necessary condition in order to be able to employ the standard inverse Higgs phenomenon and additional conditions on the structure constants must be met. This was touched upon by McArthur in \cite{mcarthur} and in the following we give a complimentary discussion with some important differences.

It turns out that (\ref{structureproject}) is a necessary condition for \textit{all} parametrisations of the coset element, however the additional conditions are heavily parametrisation dependent. We illustrate this below with three examples where for clarity we will assume that the covariant derivative forms an irreducible representation of the subgroup $H$ such that the inverse Higgs constraint comes from setting (\ref{covderiv}) to zero, rather than a projection. Now given that the vielbein is non-zero this is equivalent to setting the Maurer-Cartan component $ (\o^a)_\n$ to zero. In this case we require a commutator of the form
\begin{equation} \label{IHcomm}
[P_{\m},T_{m}] \supset T_{a} \,,
\end{equation}
if $\f^{m}$ is to appear linearly in $ (\o^a)_\n$. Since $(\o^a)_\n$ is linear in derivatives, in order to be able to algebraically solve for $\f^m$ no $\partial_\m \f^m$ terms are allowed to be present.   

The first coset parametrisation one might consider is 
\begin{equation}
\g = e^{x^{\m}P_{\m} + \f^{A}T_{A}} = e^{x^{\m}P_{\m} + \f^{a}T_{a} + \f^{m}T_{m}} \,,
\end{equation}
where all generators appear in a single exponential. This turns out to be a bad choice, not least because the resulting non-linear realisation will have explict coordinate dependence and translations act in a non-standard way on the coset coordinates, but also the condition (\ref{IHcomm}) guarantees that $(\o^a)_\m$ contains $\partial_\m \f^m$ terms. Explicitly we have
\begin{equation}
(\o^a)_\m \supset -\frac{1}{2}f_{\n m}^{a} x^{\n} \partial_{\m} \f^{m} \,,
\end{equation}
and therefore one cannot employ the standard inverse Higgs constraint to eliminate the inessential Goldstone field $\phi^{m}$ algebraically. This example already clearly demonstrates that ones choice of the coset parametrisation is important with regards to the existence of inverse Higgs constraints.

The next obvious choice is the following standard parametrisation
\begin{align} \label{standardcoset}
\g = e^{x^\m P_\m}e^{\f^A T_A} = e^{x^\m P_\m}e^{ \f^{a}T_{a} + \f^{m}T_{m}} \,,
\end{align}
as used in the original papers \cite{spacetime1,spacetime2}. Unlike the previous example this choice ensures that the non-linear realisations have no explict coordinate dependence. By calculating the Maurer-Cartan form for this coset element it follows that
\begin{align}
(\o^a)_\m =& \f^{A} f_{\m A}^{a} + \partial_{\m} \f^{a} - \tfrac{1}{2!} \f^{A}(\f^{B} f_{\m A}^{I}  f_{B I}^{a} + \partial_{\m} \f^{B} f_{AB}^{a}) + \tfrac{1}{3!} \f^{A} \f^{B}(\f^{C} f_{\m A}^{I}  f_{B I}^{J} f_{CJ}^{a} + \partial_{\m} \f^{C} f_{BC}^{I} f_{AI}^{a}) \notag \\
& 
- \tfrac{1}{4!} \f^{A} \f^{B} \f^{C}(\f^{D} f_{\m A}^{I}  f_{B I}^{J} f_{CJ}^{K} f_{DK}^{a} + \partial_{\m} \f^{D} f_{CD}^{I} f_{AI}^{J} f_{BJ}^{a}) \ldots \,,
\end{align}
and therefore we require the sequence
\begin{equation} \label{standardconditions}
f_{A m}^a, \quad f_{B m}^{I}f_{A I}^a, \quad f_{Cm}^{I} f_{AI}^{J} f_{BJ}^{a}, \quad\ldots
\end{equation} 
to vanish for $(\o^a)_\m$ to be independent of $\partial_\m \f^m$.

Another possibility is to further split the broken generators into three separate exponentials like so 
\begin{align} \label{splitcoset}
\g = e^{x^\m P_\m}e^{\f^{a} T_{a}} e^{\f^m T_m}.
\end{align}
Computing the Maurer-Cartan form for this coset element, it follows that
\begin{align}
(\o^a)_\m =& \ldots -\tfrac{1}{2!} (\f^{m} \partial_{\m} \f^{n} f_{mn}^{a} + \ldots) + \tfrac{1}{3!}(\f^{m} \f^{q} \partial_{\m}\f^{n} f_{mn}^{I} f_{qI}^{a} + \ldots) + \notag \\
& - \tfrac{1}{4!} (\f^{m} \f^{q} \f^{r} \partial_{\m}\f^{n} f_{mn}^{I} f_{qI}^{J} f_{rJ}^{a} + \ldots) + \ldots \,,
\end{align}
where for brevity we have concentrated only on the $\partial_{\m}\f^{n}$ dependence. In this case we therefore require the sequence 
\begin{equation} \label{splitconditions}
f_{mn}^a,\quad f_{mn}^{I}f_{q I}^a, \quad f_{m n}^{I}f_{q I}^{J}f_{r J}^a, \quad \ldots
\end{equation} 
to vanish for $(\o^a)_\m$ to be independent of $\partial_\m \f^m$. It is clear that the two sets of conditions (\ref{standardconditions}) and (\ref{splitconditions}) are different but interestingly the later conditions are the least stringent. In fact, out of all the possible parametrisations of the coset element, this parametrisation leads to the least stringent conditions on the structure constants and is therefore the best parametrisation to use if one wishes to find a non-linear realisation on a reduced set of fields. 

We illustrate these points below with an example which also emphasises the importance of considering the conditions on the structure conditions beyond linear order as we have done here.

\bigskip

\noindent
{\bf Example:} Consider the spontaneous breaking of the $d$-dimensional Poincar\'{e} group down to its $(d-1)$-dimensional subgroup i.e. the coset space
\begin{equation}
\text{ISO}(d-1,1)/\text{SO}(d-2,1).
\end{equation}
The $d$-dimensional Poincar\'{e} algebra has the following non-vanishing commutators
\begin{align}
&[M_{AB},P_C] = \eta_{AC}P_B - \eta_{BC}P_A \,, \quad
[M_{AB},M_{CD}] = \eta_{AC}M_{BD} - \eta_{BC}M_{AD} + \eta_{BD}M_{AC} - \eta_{AD}M_{BC} \, ,
\end{align}
where the indices $A,B,C,\ldots$ are $d$-dimensional spacetime indices and we use the Minkowski metric $\eta_{AB} = (-,+,+,\cdots)$. We initially use the standard parametrisation (\ref{standardcoset}) for the coset element such that
\begin{equation}
\g =  e^{x^\m P_\m}e^{\pi P_{d} + \Omega^{\m}M_{\m d}} \,,
\end{equation}
where $\m = 0,1,\ldots d-1$ and $P_{d}$, $M_{\m d}$ are respectively the generators of broken translations and rotations. The commutator
\begin{align}
[P_{\m},M_{\n d}] = -\eta_{\m\n} P_{d} \,,
\end{align}
informs us that $\Omega^{\m}$ appears linearly in the Maurer-Cartan component associated with $P_{d}$, $(\o_{P_{d}})_{\m}$. The covariant derivative associated with $P_{d}$ is indeed irreducible so in principle the inverse Higgs constraint would come from setting $(\o_{P_{d}})_{\m} = 0$. However the structure constants do not satisfy the series of constraints (\ref{standardconditions}) and so this Maurer-Cartan component will contain derivatives of $\Omega^{\m}$ so we cannot set it to zero to solve for $\Omega^{\m}$ as a function of $\pi$ and $\partial^{\m}\pi$. Indeed the would-be inverse Higgs constraint is
\begin{equation}
\frac{\sin \sqrt{\Omega^2}}{\sqrt{\Omega^2}} \partial_{\mu}\pi - \frac{\sin \sqrt{\Omega^2}}{\sqrt{\Omega^2}} \Omega_{\mu} + \frac{\pi}{\sqrt{\Omega^2}\Omega^{2}}(\sqrt{\Omega^2} - \sin \sqrt{\Omega^2}) \Omega_{\n}\partial_{\m} \Omega^{\n} = 0.
\end{equation}
As can be seen from expanding $ \sin \sqrt{\Omega^2}$, the leading order derivative piece is of the form $\sim \pi \Omega_{\n}\partial_{\m} \Omega^{\n}$ indicating that the leading order condition on the structure constants is satisfied but the next to leading order one i.e. $f_{Bm}^{I}f_{AI}^a = 0$ is not. So for this particular parametrisation of the coset element it is not possible to non-linearly realise the $d$-dimensional Poincar\'{e} group with a reduced set of fields.

If we instead employ the split parametrisation \eqref{splitcoset} then an inverse Higgs constraint does exist. Now the coset element reads
\begin{equation}
\g =  e^{x^\m P_\m}e^{\pi P_{d}}e^{\Omega^{\m}M_{\m d}} \,,
\end{equation}
and by setting the Maurer-Cartan component along the broken generator $P_{d}$ to zero we arrive at the inverse Higgs constraint
\begin{equation}
\cos \sqrt{\Omega^2} \partial_{\mu}\pi - \frac{\sin \sqrt{\Omega^{2}}}{\sqrt{\Omega^{2}}}\Omega_{\m} = 0 \,,
\end{equation}
which has a linear piece, and is fully algebraic, in $\Omega^{\m}$ so we can use this equation to eliminate all dependence of the Maurer-Cartan form on $\Omega^{\m}$ in favour of the essential Goldstone $\pi$. The resulting non-linear realisation corresponds to the DBI galileons \cite{reunited} in $d-1$ dimensions with the leading order term simply the scalar sector of the $(d-1)$-dimensional DBI action. We refer the reader to \cite{wess} for more details.

\subsection{Generalised inverse Higgs constraints}

In some cases it is possible to impose a ``generalised"  inverse Higgs constraint, i.e.~another way of eliminating the inessential Goldstone without spoiling the non-linear realisation. As we mentioned in the introduction, this could be an equation of motion if the inessential Goldstone is an auxilliary field, or it could arise from integrating out the inessential Goldstone at low energies. In this subsection we show that, as long as one satisfies the series of conditions for the inverse Higgs constraints discussed above, the structure of the effective field theory which non-linearly realises the broken symmetry group in terms of only the essential Goldstone is always the same. In particular, it is independent of whether  the standard inverse Higgs constraint or a generalised one is imposed; the only possible differences lie in those coupling constants which are not fixed by symmetry. 

To see this we first note that the most general transformation rules for the coordinates and the fields are
\begin{align}
g \cdot x^{\m} &= x^{\m} + h^{\m}(c_{i},x,\f,\xi)\,, \quad 
g \cdot \f^{a} = \f^{a} + f^{a}(c_{i},x,\f,\xi) \,, \quad 
g \cdot \xi^{m}  = \xi^{m} + g^{m}(c_{i},x,\f,\xi) \,, \label{trans3}
\end{align}
where $c_{i}$ are the symmetry parameters and by virtue of the Baker-Campbell-Hausdorff (BCH) formula the functions $h,f,g$ admit a standard Taylor expansion around the origin. For each function the leading order piece is bi-linear in the coset coordinates and the symmetry parameters.

Now we assume that the conditions outlined in the previous section to use the inverse Higgs phenomenon have been met. That is, the covariant derivative $\nabla_{\m} \f^{a}$ is irreducible and the Maurer-Cartan component $(\o^a)_\m$ does not depend on $\partial_\m \xi^{m}$ such that if we wanted to use the standard inverse Higgs constraint we could set $(\o^a)_\m = 0$. We therefore have $\o^a = (\o^a)_\m (\f,\partial\f,\xi) dx^{\m} $ and since this object transforms covariantly we have
\begin{align}
g \cdot (\o^a)_\m (\f,\partial\f,\xi) dx^{\m} &=(\o^a)_\m(g \cdot \f, g \cdot \partial\f,g \cdot \xi) g \cdot dx^{\m} = D^{a}_{b}(h)(\o^b)_\m  (\f,\partial\f,\xi) dx^{\m} .
\end{align} 
It follows that the product $(\o^a)_\m(g \cdot \f, g \cdot \partial\f,g \cdot \xi) g \cdot dx^{\m}$ must be independent of $\partial_{\m}\xi^{m}$.
Now given the transformation rule (\ref{trans3}) we have $ g \cdot dx^{\m} = (\delta^{\m}_{\n} + \partial_{\n}h^{\m} + h^{\m}_{\f^{a}}\partial_{\n}\f^{a}  +  h^{\m}_{\xi^{m}}\partial_{\n}\xi^{m}) dx^{\n}$ and therefore consistency of the coset construction requires
\begin{eqnarray} \label{gihcondition2}
 (\o^a)_\m(g \cdot \f, g \cdot \partial\f,g \cdot \xi)_{\partial_{\n}\xi^{m}}(\delta^{\m}_{\a} + \partial_{\a}h^{\m} + h^{\m}_{\f^{a}}\partial_{\a}\f^{a}  +  h^{\m}_{\xi^{m}}\partial_{\a}\xi^{m}) dx^{\a} \nonumber \\ +  (\o^a)_\m(g \cdot \f, g \cdot \partial\f,g \cdot \xi) h^{\m}_{\xi^{m}} dx^{\n}= 0.
\end{eqnarray}
Now we wish to derive the conditions on the symmetry transformations such that (\ref{gihcondition2}) is solved. This solution must hold for all symmetry parameters and all field values so we can perform an order by order analysis. At lowest order in fields and parameters equation (\ref{gihcondition2}) reduces to
\begin{equation}
 (\o^a)_\m(g \cdot \f, g \cdot \partial\f,g \cdot \xi)_{\partial_{\n}\xi^{m}} = 0 \,,
\end{equation}
due to the $\delta^{\m}_{\n} dx^{\n}$ piece of $g \cdot dx^{\m}$ and since $g \cdot \f$ and $g \cdot \xi$ are already independent of $\partial_{\m}\xi^{m}$ this is equivalent to 
\begin{equation} \label{gihcondition3}
 (g \cdot \partial_{\m}\f^{a})_{\partial_{\n}\xi^{m}} = 0 \,.
\end{equation}
It follows from (\ref{trans3}) that
\begin{equation}
g \cdot \partial_{\m} \f^{a} = (\delta^{\n}_{\m} + \partial_{\m}h^{\n} + h^{\n}_{\f^{a}} \partial_{\m}\f^{a} + h^{\n}_{\xi^{m}} \partial_{\m}\xi^{m})^{-1}(\partial_{\n}\f^{a} + \partial_{\n}f^{a} + f^{a}_{\f^{b}} \partial_{\n}\f^{b} + f^{a}_{\xi^{m}} \partial_{\n}\xi^{m}) \,,
\end{equation}	
and then by computing  $(g \cdot \partial_{\m}\f^{a})_{\partial_{\n}\xi^{m}}$ to lowest order in fields it is clear that the only solution to (\ref{gihcondition3}) is
\begin{align}
f^{a}_{\xi^{m}} =0 \,, \quad h_{\xi^{m}}^{\m} = 0 \,.
\end{align}
Using an iterative argument we can then conclude that to \textit{all} orders in fields and symmetry parameters $f^{a}$ and $h^{\m}$ must be independent of $\xi^{m}$. 

Therefore if one constructs the non-linear realisation with only the true Goldstone from the bottom up using the first two symmetry transformations of (\ref{trans3}), the structure of the effective field theory does not depend on how one eliminates the inessential Goldstone since it drops out of the symmetry transformations. This equivalence has been discussed in the literature before, for example \cite{brauner, wheel}, but to our knowledge this is the first time it has been proven to be true. Below we illustrate this equivalence with an informative example\footnote{We thank Joaquim Gomis for drawing our attention to this example.}.

\bigskip

\noindent {\bf Example:} Consider the spontaneous breaking of the conformal group in one dimension corresponding to the coset space
\begin{equation}
SO(1,2)/\mathds{1}.
\end{equation}
The generators are $P,D$ and $K$ and the algebra is
\begin{equation}
[P,D] = P, ~~~~~~~ [D,K] = K, ~~~~~~~ [P,K] = -2D.
\end{equation}
Given our discussion in the previous subsection, we take the coset element as
\begin{equation}
\g = e^{t P}e^{\phi D}e^{\psi K} \,,
\end{equation} 
to maximise our chances of finding a standard inverse Higgs constraint. One can straightforwardly compute the corresponding Maurer-Cartan form which is given by
\begin{equation}
\g^{-1}d \g =  e^{\phi} dt P + (d \phi - 2 \psi e^{\phi} dt) D + (d\psi + \psi d\phi - \psi^{2}e^{\phi}dt)K \,.
\end{equation}
Now consider the following invariant action
\begin{equation} \label{conformal1D}
S = \int   e^{\phi} dt  (g_{1}  - 2g_{2}\psi  + g_{3}( e^{-\phi}\psi \dot{\phi} - \psi^{2})) \,,
\end{equation} 
where we have taken a linear sum of the Maurer-Cartan components each with a coupling constant $g_{i}$ and dropped total derivatives.

It is clear that one can set the Maurer-Cartan component associated with the generator $D$ to zero such that we can solve for $\psi$ in terms of $\phi$ and its derivatives. Doing so yields
\begin{equation} \label{conformal1}
\psi = \frac{1}{2}e^{-\phi} \dot{\phi}.
\end{equation}
Imposing this constraint on our invariant action we arrive at (up to total derivatives)
\begin{equation}
S = \int  e^{\phi}dt \left(g_{1} + \frac{g_{3}}{4}e^{-2\phi}\dot{\phi}^{2} \right).
\end{equation} 
However, given that (\ref{conformal1D}) is algebraic in the field $\psi$ we can also eliminate it via its equation of motion yielding the new constraint
\begin{equation}
\psi = -\frac{g_{2}}{g_{3}} + \frac{1}{2}e^{-\phi} \dot{\phi} \,,
\end{equation}
which differs by a constant from the standard inverse Higgs constraint. Upon imposing this constraint on our invariant action we arrive at (again dropping total derivatives)
\begin{equation}
S = \int e^{\phi}dt  \left(g_{1} + \frac{g_{2}^{2}}{g_{3}} + \frac{g_{3}}{4}e^{-2\phi}\dot{\phi}^{2} \right) \,.
\end{equation} 
We see that  imposing the two different constraints indeed yields the same effective field theory but with different coupling constants.

\section{Mapping non-linear realisations} \label{mapping}

In this section we examine how non-linear realisations obtained from different coset parametrisations are related, both before and after the inessential Goldstones have been eliminated. We will only discuss this relation in the \textit{absence} of external sources and we refer the reader to \cite{duality2} for a discussion of the subtleties which arise there. During our analysis, we will encounter various types of transformations relating the different coset constructions. Let us first discuss these very briefly to set the stage for this section. 

Prior to inverse Higgs, the natural transformations are standard redefinitions of the coset coordinates. In the purely internal case these are simply field redefinitions $\bar{\f}^A = \bar{\f}^A(\f)$ whereas for spacetime symmetries they are so-called \textit{point transformations} which mix both fields and the spacetime coordinates: $\bar{x}^\m = \bar{x}^\m(x,\f)$ and $\bar{\f}^A = \bar{\f}^A(x,\f)$.

Post inverse Higgs, the natural transformations also involve derivatives, since the inessential Goldstones have been eliminated in favour of the essential ones and their derivatives. Here we encounter so-called \textit{contact transformations} and their generalisation \textit{extended contact transformations}. Starting with the former, an $n$-th order contact transformation is any transformation of the form
\begin{align}  
\bar{x}^\m = \bar{x}^\m(x,\f,...,\partial^n \f),\qquad \bar{\f}^a = \bar{\f}^a(x,\f,...,\partial^n \f), \label{contacttrans}
\end{align}
that maps the following sets onto each other:
\begin{align} 
(x^\m,\f^a, \partial \f^a, \ldots, \partial^n \f^a) \leftrightarrow (\bar{x}^\m,\bar{\f}^a,\bar{\partial}\bar{\f}^a,\ldots, \bar{\partial}^n \bar{\f}^a). \label{contactrequirement}
\end{align}  
Interestingly, non-trivial contact transformations only exist when $\f^{a}$ has a single component. Moreover, they are always first order i.e. $n=1$ \cite{LieBacklund}. Well known contact transformations are the AdS-conformal mapping considered in \cite{equivalence,mapping} and the galileon duality\footnote{See \cite{apples} for a very recent discussion on this duality in the context of UV properties of galileons \cite{uv}.} \cite{duality,duality1} that follows as a limiting case. 

Since one often deals with multiple component fields, we are naturally led to \textit{extended contact transformations}\cite{remko}. An $n$-th order extended contact transformation is a transformation of the form (\ref{contacttrans}), but without the additional requirement (\ref{contactrequirement}). Non-trivial transformations of this type \textit{do} exist for any order $n$. These are the most general local redefinitions one can perform, i.e.~they are the local subset of the Lie-B\"{a}cklund transformations \cite{LieBacklund}. As such, they include as special cases all of the previously mentioned transformations, i.e.~field redefinitions, point- and contact transformations. 

\subsection{\textit{Prior} to inverse Higgs: point transformations}
As already noted, for a given coset space one can parametrise the coset element in many different ways. For some particular basis for the broken generators $T_A$ we can put all the generators in a single exponential, every generator in a separate exponential, or anything inbetween. In addition, the order of the exponentials is freely specifiable. To be more precise, one can consider any partition $A = (a_1,...,a_k)$ and subsequently parametrise the coset element as
\begin{align}  
\g = e^{\f^{a_1}T_{a_1}} \cdots 	e^{\f^{a_k}T_{a_k}} \,, \label{unbarredparam}
\end{align}  
where we have temporarily include $P_\m$ in $T_A$ for notational convenience.

A further freedom lies in the choice of algebra basis for the broken generators. That is, one can consider an alternative basis $\bar{T}_A$ invertibly related to the original one by
\begin{align}  
\bar{T}_A = c_A^B T_B + c_A^i T_i, \qquad \qquad\det (c_A^B) \neq 0. \label{barredbasis}
\end{align}  
Again in this basis, one can pick any partition $A = (a'_1,...,a'_{l})$ and use the corresponding parametrisation
\begin{align} 
\bar{\g} = e^{\bar{\f}^{a'_1}\bar{T}_{a'_1}} \cdots 	e^{\bar{\f}^{a'_l}\bar{T}_{a'_l}}. \label{barredparam}
\end{align}
A physically interesting example of such different bases arises in the context of the conformal group and branes in AdS space and will be discussed in section \ref{conformal}.

Given any two bases related by (\ref{barredbasis}) and any two corresponding arbitrary partitions, it follows from the BCH formula that there exists a (locally) invertible redefinition of the coset coordinates relating the corresponding parametrisations. That is, one has 
\begin{align}  
\g = e^{\f^{a_1}T_{a_1}} \cdots 	e^{\f^{a_k}T_{a_k}}  =  e^{\bar{\f}^{a'_1}\bar{T}_{a'_1}} \cdots 	e^{\bar{\f}^{a'_l}\bar{T}_{a'_l}}\cdot e^{\bar{\f}^iT_i} = \bar{\g} h,  \label{relationparams}
\end{align}
where  
\begin{align} 
\bar{\f}^A &= \bar{\f}^A(\f^B) = (c^{-1})^A_B\f^B + \textup{ terms higher order in coset coordinates,} \label{pointtrans} \\
\bar{\f}^i &= \bar{\f}^i(\f^B) =  -c^i_A (c^{-1})^{A}_{B}\f^B + \textup{ terms higher order in coset coordinates,}   \label{pointtrans1}
\end{align}
and invertibility of (\ref{pointtrans}) is guaranteed by the presence of the linear term. The exact form of the resulting mapping can be highly non-trivial on account of the BCH formula.

The relation (\ref{relationparams}) induces an equivalence of the corresponding non-linear realisations. If their transformation laws are $g \cdot (x^\m,\f^A)$ and $g\cdot (\bar{x}^\m,\bar{\f}^A)$ then 
\begin{align} 
\bar{x}^\m(g \cdot (x,\f)) = g\cdot \bar{x}^\m, \qquad \bar{\f}^A(g \cdot (x,\f)) =  g\cdot\bar{\f}^A. \label{transformationsymmetry}
\end{align}  
Thus starting from any action $S[x,\f]$ which is invariant under $g\cdot (x^\m,\f^A)$, one can obtain an equivalent barred action $\bar{S}[\bar{x},\bar{\f}]$ invariant under $g\cdot (\bar{x}^\m,\bar{\f}^A)$ by performing the point transformation (\ref{pointtrans}). In other words
\begin{align} 
S = \int L(x,\f)\sqrt{-g}d^{d}x \equiv  \int \bar{L}(\bar{x},\bar{\f}) \sqrt{-\bar{g}}d^{d}\bar{x} = \bar{S},
\end{align} 
where we have defined
\begin{align} 
\bar{L}(\bar{x},\bar{\f}) \equiv L(x,\f)\frac{\sqrt{-g}}{\sqrt{-\bar{g}}}\det\Big(\frac{dx^{\m}}{d\bar{x}^{\n}}\Big). \label{Lagrangianpoint}
\end{align} 
Therefore using either of the parametrisations results in equivalent physical theories. 

Universality of the coset construction, in the sense that \textit{any} non-linear realisation can be brought back to a specific coset form, has only been proven for compact, connected and semi-simple Lie groups. For more general internal symmetries and spacetime symmetries, there is no proof of universality and therefore it is not clear that \textit{any} non-linear realisation can be brought back to the coset form, even prior to inverse Higgs. There are, however, examples where it is possible.

An interesting example relates to supersymmetry, corresponding to the spontaneous breaking of super-Poincar\'{e} to the Poincar\'{e} group. The corresponding coset element contains a fermion field, the Goldstino, but no inessential Goldstone modes and hence there are no inverse Higgs constraints. However, other methods can be used to arrive at a non-linear realisation of supersymmetry, for example, by imposing a supersymmetric constraint on a linear supermultiplet (see e.g.~\cite{Rocek, Casalbuoni, Komargodski}). In this case an explicit point transformation relating this non-linear realisation to the one coming from the coset construction of \cite{VA} has been constructed \cite{Kuzenko}.

\subsection{\textit{Post} inverse Higgs: extended contact transformations}	
We will now examine whether the equivalence between realisations is maintained after eliminating the inessential Goldstones. We again consider parametrisations (\ref{unbarredparam}) and (\ref{barredparam}) and assume that for both we can consistently employ the standard inverse Higgs mechanism to remove the inessential fields. For clarity we again focus on a single inverse Higgs constraint:
\begin{eqnarray}
IHC: \qquad \nabla_\m \f^a|^{m} = 0 \, \qquad \Leftrightarrow \qquad &\f^m - \mathcal{F}^m (\f^a,\partial_\m \f^a) = 0\,, \\
\overline{IHC}: \qquad\bar{\nabla}_\m\bar{\f}^a|^{m} = 0 \, \qquad \Leftrightarrow \qquad &\bar{\f}^m - \bar{\mathcal{F}}^m (\bar{\f}^a, \bar{\partial}_\m \bar{\f}^a) = 0.
\end{eqnarray}
If the point transformation relating the two sets of coset coordinates \textit{prior} to imposing inverse Higgs constraints is to induce an invertible transformation relating the essential coordinates to each other \textit{post} inverse Higgs, one must demand compatibility in the following sense
\begin{align} 
\bar{\f}^A(\f|_{IHC}) &= \bar{\f}^A|_{\overline{IHC}} \,. \label{inducedtransformation}  
\end{align}
This is precisely the case when 
\begin{align}  
\nabla_\m \f^a|^{m} = 0 \qquad \Leftrightarrow \qquad \bar{\nabla}_\m\bar{\f}^a|^{m} = 0 \label{compIH},
\end{align} 
i.e. when the two inverse Higgs constraints imply each other based on the point transformation relating the coset elements (see \cite{duality1,specialgeometry} for a discussion related to Galileons). If this is indeed the case then the induced transformation relating the spacetime coordinates and the essential Goldstones is simply the point transformation evaluated on the inverse Higgs constraints
\begin{align}   \label{contactfrompoint}
\bar{x}^\m = \bar{x}^\m(x,\f,\mathcal{F}(\f,\partial \f)), \qquad \bar{\f}^a = \bar{\f}^a(x,\f,\mathcal{F}(\f,\partial \f)).
\end{align} 
The result is an invertible first order\footnote{ We note that this potentially generalises to a $n$-th order extended contact transformation when there are $n$ inverse Higgs constraints.} extended contact transformation which reduces to a standard contact transformation when $\f^{a}$ contains a single component. Its invertibility is guaranteed by that of the point transformation.

Due to the compatibility of the inverse Higgs conditions and the point transformation, the transformation rules for the essential Goldstones (and spacetime coordinates) are mapped onto each other under the extended contact transformation. Again this ensures physical equivalence of the post inverse Higgs non-linear realisations. In particular, two equivalent Lagrangians prior to inverse Higgs remain equivalent post inverse Higgs. It is interesting to note, however, that due to the derivative nature of the extended contact transformations, the order of a Lagrangian is generically not maintained.

On the other hand if (\ref{compIH}) is not satisfied, it is far from clear if equivalence is maintained post inverse Higgs. What we can say for sure is that if an invertible mapping does exist, it does not directly follow from the point transformation relating the coset elements. This is a somewhat surpising possibility but it is very easy to find situations where it occurs. To see this, consider two Maurer-Cartan forms prior to imposing inverse Higgs constraints where the corresponding coset elements are related by (\ref{relationparams}) but we restrict to the case where $\bar{T}_{A} = c^{B}_{A}T_{B} + c^{i}_{A}T_{i} + c^{\m}_{A}P_{\m}$. Obviously here we do not combine $P_{\m}$ and $T_{A}$. The Maurer-Cartan forms are related by 
\begin{align} \label{MCmapped}
\g^{-1}d\g= h^{-1}(\bar{\g}^{-1}d\bar{\g}) h + h^{-1}dh \,,
\end{align}
or in terms of their components we have
\begin{align} 
\o^\m &= D(h^{-1})^\m_\n \bar{\o}^\n + c_b^\m D(h^{-1})^b_c \bar{\o}^c  + c_m^\m D(h^{-1})^m_n \bar{\o}^n, \notag \\
\o^A  &= c_b^A D(h^{-1})^b_c \bar{\o}^c  + c_m^A D(h^{-1})^m_n \bar{\o}^n, \notag \\
\o^i  &= D(h^{-1})^i_j \bar{\o}^j +  c_b^i D(h^{-1})^b_c \bar{\o}^c  + c_m^i D(h^{-1})^m_n \bar{\o}^n + (h^{-1}dh)^i.  \label{MCmapped1}
\end{align} 
For simplicitly let us assume that the covariant derivatives which lead to the inverse Higgs constraints are irreducible such that the unbarred inverse Higgs conditions are $\o^a = 0$ and the barred ones are $\bar{\o}^a = 0$. It follows from (\ref{MCmapped}) that in general we have
\begin{align}  \label{nonmappedih}
\bar{\o}^a = 0 \,, \qquad \nLeftrightarrow \qquad \o^a = c^a_b D(h^{-1})^b_c \bar{\o}^c + c^a_m D(h^{-1})^m_n \bar{\o}^n = 0 \,,
\end{align} 
since in general $\bar{\o}^n \neq 0$ on the inverse Higgs solutions. Here the inverse Higgs constraints are not mapped onto each other under the point transformation and therefore the point transformation does not induce a transformation relating the two non-linear realisations constructed from only the essential Goldstones. 

We also note that if one considers two parametrisations with the \textit{same} basis of broken generators, i.e. when $c^A_B = \delta^A_B$ and thus $c^a_m = 0$, one finds 
\begin{align} 
\bar{\o}^a = 0 \,, \qquad \Leftrightarrow \qquad \o^a = D(h^{-1})^a_b \bar{\o}^b = 0 \,,
\end{align}
such that the inverse Higgs constraints are indeed mapped. It follows that in this case the equivalence between the non-linear realisation is guaranteed to be maintained even after the inessential Goldstones have been eliminated.

Below we show that the non-mapping of the inverse Higgs constraints can indeed occur but does not necessarily imply inequivalence of the two non-linear realisations.

\bigskip

\noindent
{\bf Example:}
Consider spontaneous breaking of the Poincar\'{e} group in two dimensions i.e. the coset space
\begin{equation} 
ISO(1,1)/\mathds{1}.
\end{equation}
We work in two different bases for the algebra, the first with generators $P_{0}, P_{1}$ and $M$, and the second with $\bar{P}_{0} = P_{0}, \bar{P}_{1} =  P_{1}$ and $\bar{M} = M +\alpha P_{1}$. Since the generators of translations commute with each other the commutators are the same in each basis and are given by	
\begin{equation}
[P_{0},M] = P_{1}, \quad [P_{0},\bar{M}] = P_{1} \,, \quad  [P_{1},M] = P_{0} \,, \quad[P_{1},\bar{M}] = P_{0}.
\end{equation}
We parametrise the two coset elements as
\begin{equation}
\gamma = e^{tP_{0}}e^{\pi P_{1}}e^{\Omega M}, \qquad 
\bar{\gamma} = e^{\bar{t}P_{0}}e^{\bar{\pi} P_{1}}e^{\bar{\Omega} \bar{M}},
\end{equation}	
yielding the two Maurer-Cartan forms
\begin{align}
\gamma^{-1}d \gamma & = P_{0}(\cosh \Omega dt + \sinh \Omega d\pi) + P_{1}(\sinh \Omega dt + \cosh \Omega d\pi)  + M d \Omega, \notag \\
\bar{\gamma}^{-1}d \bar{\gamma} & = P_{0}(\cosh \bar{\Omega} d\bar{t} + \sinh \bar{\Omega} d\bar{\pi}) + P_{1}(\sinh \bar{\Omega} d\bar{t} + \cosh \bar{\Omega} d\bar{\pi})  + \bar{M} d \bar{\Omega} \,, \label{MC2}
\end{align}
which of course have the same structure given that the commutators are the same. The point transformation which relates these two Maurer-Cartan forms is
\begin{align}
\bar{t} = t+ \alpha \cosh \Omega \,, \quad 
\bar{\pi} = \pi - \alpha \sinh \Omega \,, \quad 
\bar{\Omega} = \Omega \,, \label{point3}
\end{align}
which is extracted by equating both expressions in (\ref{MC2}). The inverse Higgs constraints in both cases come from setting the co-efficient of $P_{1}$ in the Maurer-Cartan forms to zero, due to the commutators $[P_{0},M] = P_{1}$ and $[P_{0},\bar{M}] = P_{1}$, yielding the solutions
\begin{align}
\Omega = \tanh^{-1}(-\dot{\pi}) \,, \quad
\bar{\Omega} = \tanh^{-1}(-\pi') \,,
\end{align}
where a prime denotes a derivative with respect to $\bar{t}$. Now these solutions are not mapped onto each other under the point transformations (\ref{point3}) therefore the two Maurer-Cartan forms after we impose the inverse Higgs constraints are also not mapped onto each other. This is obvious given that in the unbarred variables the co-efficient of $P_{1}$ now vanishes due to the inverse Higgs constraint while it is non-zero in the barred basis after we set $\bar{M} = M +\alpha P_{1}$.

The resulting building blocks of invariant Lagrangians are
\begin{eqnarray}
\sqrt{1 - \dot{\pi}^{2}}dt, \qquad \frac{\ddot{\pi}}{(1-\dot{\pi}^{2})^{3/2}} , \qquad \text{and} \qquad
\sqrt{1 - \bar{\pi}'^{2}}d\bar{t}, \qquad \frac{\bar{\pi}''}{(1-\bar{\pi}'^{2})^{3/2}} \,,
\end{eqnarray}
and are therefore mapped onto each other in the trivial manner $\bar{t} = t$ and $\bar{\pi} = \pi$ post inverse Higgs but this has nothing to do with how the coset elements are related. Of course any Wess-Zumino terms will also be mapped. \\

\section{Correspondence between AdS and conformal cosets} \label{conformal}

It turns out that both cases of interest discussed above, i.e. with the inverse Higgs constraints mapped or not, apply to the spontaneous breaking of the $d$-dimensional conformal group by a codimension $d-n+1$ Minkowski brane embedded in $\text{AdS}_{d+1}$. The two different bases for the algebra are the standard conformal basis and the AdS basis \cite{equivalence}. The coset space is
\begin{equation} 
SO(d,2) / (SO(n-1,1) \times SO(d-n)) \,,
\end{equation}
where the unbroken $SO(d-n)$ transformations correspond to the unbroken Lorentz transformations in the directions transverse to the brane. Whether a mapping between invariant Lagrangians which follows from the point transformation relating the coset elements exists is dependent on the codimension of the brane. It turns out that for codimension one branes there is indeed a well defined mapping of this kind, as discussed in \cite{equivalence,mapping}, but for any other codimension this is not the case as we illustrate below.

\subsection{Codimension one}

Let us begin with the codimension one case corresponding to the coset space
\begin{equation} 
SO(d,2) / SO(d-1,1).
\end{equation}
In the standard basis of the conformal algebra the non-vanishing commutators are
\begin{equation*}
\begin{aligned}[c]
&[P_A,D] = P_A\, \\
&[K_A,D] = -K_A\,\\
&[P_A,K_B] = 2M_{AB} + 2\eta_{AB}D\,
\end{aligned}
\qquad
\begin{aligned}[c]
&[M_{AB},P_C] = \eta_{AC}P_B - \eta_{BC}P_A \\
&[M_{AB},K_C] = \eta_{AC}K_B - \eta_{BC}K_A \\
&[M_{AB},M_{CD}] = \eta_{AC}M_{BD} - \eta_{BC}M_{AD} + \eta_{BD}M_{AC} - \eta_{AD}M_{BC},
\end{aligned}
\end{equation*}
where again we use $A,B,C,\ldots$ for $d$-dimensional spacetime indices. The $d+1$ broken generators correspond to dilatations $D$ and special conformal transformations $K_{A}$. Given our discussion in section \ref{inverseHiggs} we parametrise the coset element as
\begin{equation}
\g = e^{x^{A}P_{A}}e^{\f D} e^{\psi^{A}K_{A}}.
\end{equation}
Now the commutator $[P_A,K_B] = 2M_{AB} + 2\eta_{AB}D$ tells us that $\psi^{A}$ appears linearly in the covariant derivative associated with $D$ and since this covariant derviative is an irrep the standard inverse Higgs constraint would come from setting the Maurer-Cartan component $\o_{D}$ to zero. Indeed the structure constants satisfy the conditions (\ref{splitconditions}) and so we can use this constraint to algebraically eliminate $\psi^{A}$ in favour of $\f$ and $\partial_{A}\f$. The resulting non-linear realisation is equivalent to building diffeomorphism invariant scalars out of the effective metric $g_{AB} = e^{2\phi}\eta_{AB}$. In four dimensions the leading terms in a derivative expansion yield the familiar Lagrangian
\begin{equation}
\mathcal{L} = -\frac{1}{2} (\partial \varphi)^{2} + \frac{\lambda}{4!}\varphi^{4},
\end{equation}
after the field redefinition $\varphi = e^{\phi}$.

The AdS basis is defined by\footnote{To compare with \cite{mapping} we are working in units where $L = 1/\sqrt{2}$.} \cite{equivalence} 
\begin{equation} \label{adsbasis}
\bar{K}_{A} = K_{A} + \frac{1}{2}P_{A},
\end{equation}
in which case the non-vanishing commutators are
\begin{equation*}
\begin{aligned}[c]
&[P_A,D] = P_A\, \\
&[\bar{K}_A,D] = -\bar{K}_A + P_A \,\\
&[P_A,\bar{K}_B] = 2M_{AB} + 2\eta_{AB}D\, \\
&[\bar{K}_A,\bar{K}_B] = 2M_{AB} \,
\end{aligned}
\qquad
\begin{aligned}[c]
&[M_{AB},P_C] = \eta_{AC}P_B - \eta_{BC}P_A \\
&[M_{AB},\bar{K}_C] = \eta_{AC}\bar{K}_B - \eta_{BC}\bar{K}_A \\
&[M_{AB},M_{CD}] = \eta_{AC}M_{BD} - \eta_{BC}M_{AD} + \eta_{BD}M_{AC} - \eta_{AD}M_{BC}.
\end{aligned}
\end{equation*}
We now parametrise the coset element as
\begin{equation} \label{adscoset}
\bar{\g} = e^{\bar{x}^{A}P_{A}}e^{\bar{\f} D} e^{\bar{\psi}^{A}\bar{K}_{A}},
\end{equation}
and again due to the commutator $[P_A,\bar{K}_B] = 2M_{AB} + 2\eta_{AB}D$, and the fact that the structure constants satisfy the conditions (\ref{splitconditions}), we can set $\bar{\o}_{D} = 0$ to leave us with a non-linear realisation constructed solely from the dilaton $\bar{\f}$. 

Now the point transformation which maps the two coset elements can be extracted by equating the two corresponding Maurer-Cartan forms. This is because whenever the unbroken generator $M_{AB}$ is generated in (\ref{adscoset}) by the BCH formula, the indices are always contracted with copies of $\bar{\psi}^{A}$ and so it drops out by symmetry. In other words the $h$ of (\ref{MCmapped}) is trivial in this case. Importantly, since (\ref{adsbasis}) does not involve the generator $D$, i.e. we have $c_{m}^{a} = 0$ when comparing to (\ref{barredbasis}), the two inverse Higgs constraints are mapped by this point transformation, see equation (\ref{nonmappedih}). A contact transformation relating the non-linear realisations constructed from the dilatons $\f$ and $\bar{\f}$ then follows by evaluating this transformation on the inverse Higgs solutions. This has been done explictly in \cite{equivalence,mapping} and we refer the reader there for more details.

\subsection{Higher codimensions}

In higher codimensions the situation is more complicated. Now consider a $d-n+1 > 1$ codimension brane where the broken generators now also include translations and Lorentz transformations. If we let $\m,\n,\ldots$ label $n$-dimensional spacetime indices and $i = n+1,\ldots,d$, then the broken generators in the conformal basis are
\begin{equation}
P_{i}, M_{\m i}, D, K_{\m}, K_{i},
\end{equation}
and similarly for the AdS basis with $K_{A} \rightarrow \bar{K}_{A}$. In general there are now $2(d-n) + 1$ Goldstone scalars and $d-n+1$ Goldstone vectors. If we parametrise the coset elements as
\begin{equation} \label{adsconformalcoset1}
\g  = e^{x^{\m}P_{\m}}e^{\pi^{i}P_{i}} e^{\f D} e^{\Omega^{\m i}M_{\m i}}  e^{\psi^{\m}K_{\m}}e^{\sigma^{i}{K_{i}}},
\end{equation}
for the conformal basis and similarly for the AdS basis again with $K_{A} \rightarrow \bar{K}_{A}$, we can use standard inverse Higgs constraints to remove all inessential Goldstones leaving us with $d-n+1$ essential Goldstone scalars. Of course here there is more than a single inverse Higgs constraint and not all of the relevant covariant derivatives are irreps. Indeed we have to perform traces to eliminate the $\sigma^{i}$ fields using the covariant derivatives associated with $\o_{M_{\m i}}$. In any case, one of the essential Goldstones is the dilaton and the other $d-n$ correspond to the broken translations and are $SO(d-n)$ invariant.

Let us concentrate on one of these inverse Higgs constraints since this will be enough to draw conclusions about possible mappings. In both bases the commutator $[M_{AB},P_C] = \eta_{AC}P_B - \eta_{BC}P_A$ tells us that the vectors $\Omega^{\m i}$ (conformal basis) and $\bar{\Omega}^{\m i}$ (AdS basis) associated with a broken Lorentz transformation $M_{\m i}$ appear linearly in the covariant derivatives associated with the broken generator $P_{i}$. Since these covariant derivatives are irreps the inverse Higgs constraints can come from setting $\o_{P_{i}} = 0$ and $\bar{\o}_{P_{i}} = 0$. With (\ref{adsconformalcoset1}) we can eliminate $\Omega^{\m i}$ and $\bar{\Omega}^{\m i}$ algebraically. 

However, now given the definition of the AdS basis (\ref{adsbasis}), these inverse Higgs constraints \textit{will not} be mapped onto each other under the point transformation which takes us from one coset element to the other unless the Maurer-Cartan component $\o_{K_{i}}$ vanishes on the inverse Higgs solutions. This is because we now have $c_{m}^{a} \neq 0$ in equation (\ref{barredbasis}). We have checked explictly for codimension two that $\o_{K_{i}} \neq 0$ on the inverse Higgs solutions and one would expect this to hold for higher codimensions too. As we discussed above this leaves us with two possibilities. Either the standard basis and the AdS basis lead to physically different non-linear realisations for the essential Goldstones when the codimension is higher than one or there is a mapping relating invariant Lagrangians which does not follow from the point transformation which maps the coset elements. 

\section{Conclusion and outlook}

Coset constructions are a powerful tool for constructing theories with non-linearly realised symmetries. For spacetime symmetries, however, they generically involve a number of inessential Goldstone modes that are dispensable for the non-linear realisation. This makes it hard to see whether all coset constructions are equivalent. Motivated by this, in this paper we have addressed two crucial aspects with regards to the inessential Goldstones. 

First of all, we have investigated different ways of eliminating the inessential Goldstones. In the literature, this often proceeds via imposing inverse Higgs constraints. In contrast to existing claims, we have demonstrated that the existence of such constraints actually requires the structure constants to satisfy a sequence of conditions as also discussed in \cite{mcarthur}. Moreover, the severity of these conditions depends on the form of the coset element, with the standard parametrisation being a suboptimal choice. Instead, the least stringent conditions arise for a coset element that consists of the largest number of exponential factors. We have also proven, under certain assumptions, that any other method of eliminating the inessential Goldstones, algebraically or otherwise, boils down to the same physics: the resulting theory can only differ in the choice of coupling constants and hence forms an identical effective field theory. 

The second issue concerns the relation between coset constructions employing different parametrisations and/or basis choices. Again the inessential Goldstones play a crucial role. Prior to the process of elimination, all coset constructions are related to each other by means of a point transformation, involving the set of essential and inessential Goldstones as well as spacetime coordinates. This naturally generalises the field redefinitions relating all coset constructions for internal symmetries. However, such a point transformation does not necessarily relate the inverse Higgs constraints for the inessential Goldstone modes. In the case where they are related, one inherits an extended contact transformation, involving the essential Goldstones, their derivatives and the spacetime coordinates, that maps the different non-linear realisations onto each other. In the case of a single inverse Higgs constraint we have
 \begin{align}
  &  \text{\rm internal symmetries:} \quad \text{\rm field redefinitions on} \quad \phi^A \,, \notag \\
   & \text{\rm spacetime symmetries:} \quad \begin{cases} \text{\rm point transformations on} \quad (x^\mu, \phi^a, \phi^m) \,, \notag \\ \quad \quad \Downarrow \text{(when IHC mapped)}  \\
    \text{\rm extended contact transformations on} \quad (x^\mu, \phi^a, \partial_\mu \phi^a ) \,. \end{cases}
 \end{align}
More generally, if we have $n$ inverse Higgs constraints then the extend contact transformation could in principle be $n$-th order.

However, we have seen that in the cases where the inverse Higgs constraints are not related by the point transformation, there is no such inherited extended contact transformation. A natural expectation would be that the resulting theories for the essential Goldstones are inequivalent. However, we have shown that this is not necessarily the case in a simple example where the inverse Higgs constraints are unrelated but the non-linear realisations result in equivalent physics. Whether the same holds for all such theories or whether this is a consequence of the simplicity of our example remains a question of high interest for future reseach. 

This crucial distinction concerning the relation of inverse Higgs constraints is beautifully illustrated in our main physical example, focussing on the relation between the conformal and the AdS basis of the $SO(2,d)$ algebra. We have considered the spontaneous breaking of this algebra as described by a $n$-dimensional Minkowski probe brane embedded in $(d+1)$-dimensional AdS space.

For both bases, our choice for the coset parametrisation (\ref{adsconformalcoset1}) was inspired by our discussion in section 3. Even though there we primarily concentrated on a single inverse Higgs constraint for clarity, the general principle still applies for multiple inverse Higgs constraints: use the largest number of exponentials which allows one to write the coset element in a $H$-invariant way, and place the inessential Goldstones to the right. However, of course for multiple inverse Higgs there is also the added subtlety of the order of the inessential Goldstones and this can play an important role. For example, if instead of (\ref{adsconformalcoset1}) we had chosen
\begin{equation} \label{adsconformalcoset}
\g  = e^{x^{\m}P_{\m}}e^{\pi^{i}P_{i}} e^{\f D} e^{\Omega^{\m i}M_{\m i}} e^{\sigma^{i}{K_{i}}}  e^{\psi^{\m}K_{\m}},
\end{equation}
where we have reversed the order of the final two exponentials, then in the AdS basis we would not have been able to remove all inessential Goldstones algebraically since $\sigma^{i}$ would appear with derivatives in the Maurer-Cartan form along the broken generator $M_{\m i}$. This is only problematic in the AdS basis since $K_{\m}$ and $K_{i}$ commute in the conformal basis.

In any case we found that whether the constraints in the conformal and the AdS basis are mapped onto each other depends on the codimension of the brane and hence on the number of essential Goldstone modes. For codimension one, i.e.~a single essential Goldstone, the solutions for the inessential Goldstone modes are mapped onto each other, as implicitly used in \cite{equivalence,mapping}. However, we find that this ceases to be true for higher codimensions which necessarily involve more essential Goldstones. This implies that in the latter case there is no straightforward extended contact transformation relating the two different coset constructions. Clearly this deserves further attention. 

Whether or not the coset construction for spacetime symmetry breaking does indeed produce universal dynamics for the essential Goldstones remains an interesting open question. Either way the inessential Goldstones will certainly play an important role.

\section*{Acknowledgements}
We would like to thank Joaquim Gomis, Garrett Goon, Antonio Padilla and Pelle Werkman for useful discussions. The authors acknowledge the Dutch funding agency ‘Netherlands Organisation for Scientific Research’ (NWO) for financial support.

\end{document}